# Observation of Tau Neutrino Interactions

DONUT Collaboration
K. Kodama[1], N. Ushida[1], C. Andreopoulos[2], N. Saoulidou[2], G. Tzanakos[2], P. Yager[3], B. Baller[4], D. Boehnlein[4], W. Freeman[4], B. Lundberg[4], J. Morfin[4], R. Rameika[4], J.C. Yun[4], J.S. Song[5], C.S. Yoon[5], S.H.Chung[5], P. Berghaus[6], M. Kubanstev[6], N.W. Reay[6], R. Sidwell[6], N. Stanton[6], S. Yoshida[6], S. Aoki[7], T. Hara[7], J.T. Rhee[8], D. Ciampa[9], C. Erickson[9], M. Graham[9], K. Heller[9], R. Rusack[9], R. Schwienhorst[9], J. Sielaff[9], J. Trammell[9], J. Wilcox[9] K. Hoshino[10], H. Jiko[10], M. Miyanishi[10], M. Komatsu[10], M. Nakamura[10], T. Nakano[10], K. Niwa[10], N. Nonaka[10], K. Okada[10], O. Sato[10], T. Akdogan[11], V. Paolone[11], C. Rosenfeld[12] A. Kulik[11,12], T. Kafka[13], W. Oliver[13], T. Patzak[13], J. Schneps[13]

[1] *Aichi University of Education, Kariya, Japan*
[2] *University of Athens, Greece*
[3] *University of California/Davis, Davis, California*
[4] *Fermilab, Batavia, Illinois 60510*
[5] *Gyeongsang University, Chinju, Korea*
[6] *Kansas State University, Manhattan, Kansas*
[7] *Kobe University, Kobe, Japan*
[8] *Kon-kuk University, Korea*
[9] *University of Minnesota, Minnesota*
[10] *Nagoya University, Nagoya 464-8602, Japan*
[11] *University of Pittsburgh, Pittsburgh, Pennsylvania 15260*
[12] *University of South Carolina, Columbia, South Carolina*
[13] *Tufts University, Medford, Massachusetts 02155*

**Abstract**

The DONUT experiment has analyzed 203 neutrino interactions recorded in nuclear emulsion targets. A decay search has found evi-dence of four tau neutrino interactions with an estimated background of 0.34 events. This number is consistent with the Standard Model expectation.

The $\nu_\tau$ was postulated to exist after the discovery of the $\tau$ lepton in 1975 [1]. Since that time, much indirect evidence has been gathered implying that the $\nu_\tau$ exists as the Standard Model third generation neutrino. However, the charged current interactions of a third neutrino have not been observed in the same manner as the interactions of the $\nu_e$ [2] and the $\nu_\mu$ [3].

The DONUT experiment (Fermilab E872) was designed to observe the charged current interactions of the $\nu_\tau$ by identifying the $\tau$ lepton as the only lepton created at the interaction vertex. At the neutrino energies in this experiment, the $\tau$ typically decays within 2 mm of its creation to a single charged daughter (86% branching fraction). Thus the signature of the $\tau$ is a track with a kink, signifying a decay characterized by a large transverse momentum. Nuclear emulsion was used to locate and resolve these decays. A charged particle spectrometer with electron and muon identification capabilities provided additional information.

The neutrino beam was created using 800 GeV protons from the Fermilab Tevatron interacting in a meter long tungsten beam dump, which was 36 m upstream from the emulsion target. Most of the neutrinos that interacted in the emulsion target originated in the decays of charmed mesons in the beam dump. The primary source of $\nu_\tau$ is the leptonic decay of a $D_S$ meson into $\tau$ and $\overline{\nu}_\tau$, and the subsequent decay of the $\tau$ to a $\nu_\tau$. All other sources of $\nu_\tau$ are estimated to have contributed an additional 15%. $(5 \pm 1)\%$ of all neutrino interactions detected in the emulsion were predicted to be from $\nu_\tau$ with the dominant uncertainty from charm production and $D_S \to \tau\nu$ branching ratio measurements[4]. The mean energies of the detected neutrino interactions were calculated to be 89 GeV, 69 GeV, and 111 GeV, for $\nu_e$, $\nu_\mu$, and $\nu_\tau$ respectively.

The other products from the proton interactions, mostly muons, were reduced in the emulsion region using magnets, concrete, iron and lead shielding, shown in Figure 1. The emulsion targets were exposed from April to September 1997. The length of exposure for each target was set by the track density from muons, with a limit of $10^5$ cm$^{-2}$ that was reached within about one month.

The detectors, downstream of the shielding, consisted of a scintillation counter veto wall, the emulsion-scintillating fiber hybrid target (emulsion modules interleaved with scintillating fiber planes), trigger hodoscopes, and the charged particle spectrometer. Up to four emulsion modules, each 7 cm thick, were mounted along the neutrino beam, separated by $\sim$ 20 cm. Two types of emulsion modules were used as neutrino targets.

The first type (called *ECC*) had a repeated structure of 1 mm thick stainless steel sheets interleaved with emulsion plates. The emulsion plates had 100 $\mu$m thick emulsion layers on each side of a 200 $\mu$m or 800 $\mu$m thick plastic base. The mass of the emulsion in the ECC target was only 8% of the total, so that the neutrino interactions most likely occured in the steel plates. The second type of emulsion target (called *bulk*) was constructed from emulsion plates with 350 $\mu$m thick emulsion layers on both sides of a 90 $\mu$m thick plastic base. The three designs are shown schematically in Fig. 2.

Each target module consisted of either: ECC type only, bulk only, or a combination of ECC type for the upstream part and bulk for the downstream part. Two pure ECC modules, one pure bulk module and four ECC/bulk hybrid modules were exposed in the neutrino beam. The pure ECC modules had a mass of 104 kg, and were 3.0 radiation lengths and 0.23 interaction lengths (for $\pi$+N interactions). The pure bulk module had a mass of 60 kg, and was 2.0 radiation lengths and 0.13 interaction lengths. The typical ECC/bulk hybrid module was 70 kg, 2.0 radiation lengths and 0.16 interaction lengths. All of the emulsion and steel sheets were 50 cm $\times$ 50 cm oriented perpendicular to the beamline.

Distributed between the emulsion modules were 44 planes of scintillating fiber tracker (SFT) of 0.5 mm diameter read out by 6 image intensifiers [8]. The SFT in the hybrid target provided the information to locate the interaction vertex within the emulsion module. The reconstructed tracks were used to predict a vertex position with a typical precision of 1 mm in the transverse coordinates and 7 mm along the beam. The spectrometer provided the time-stamped electronic information about the event including the trigger, and lepton identification. The spectrometer readout was triggered by three hodoscopes placed in the emulsion module region consistent with more than one charged track, and no signal from the upstream veto wall. A total of $4.0 \times 10^6$ triggers were recorded from $3.54 \times 10^{17}$ protons in the tungsten beam dump. Tracks in the SFT were projected downstream to those in drift chambers for possible matches. For the tracks that were connected, momenta were measured with a 1.85 m $\times$ 1.45 m aperture magnet with a strength of 0.76 T-m. In the analysis, the spectrometer was used mainly to measure the momenta of muons. The momenta of other tracks in selected interactions were determined by measuring displacements due to multiple scattering in the emulsion modules. Muons were identified with an array of proportional tubes between 3 walls of iron (from upstream) 0.4 m, 0.9 m, and 0.9 m thick. Electrons were tagged by electron-pair creation in the emulsion, shower development in the SFT, or the lead glass calorimeter,

Triggers were selected for an emulsion search by requiring that a vertex reconstructed from tracks be within an emulsion target. In addition, the event trigger timing between two counters was required to be within 10 ns, and show no evidence of nearby showering from muon interactions. Remaining events with a visible energy more than 2 GeV and a vertex in the emulsion volume were classified as neutrino interaction candidates. A total of 898 such candidates were identified from the spectrometer information compared to the expected number of $1040 \pm 200$.

Of the 898 reconstructed neutrino interaction candidates, 698 had a predicted interaction vertex within a fiducial emulsion volume that included 80% of the total target mass. There were additional requirements of event topology and predicted vertex precision to make the sample amenable to automated scanning. This reduced the number of events that were scanned to 499.

The first step in locating an interaction vertex was scanning a small area in the relevant emulsion plates. The tracks recorded in a plate with angles $\leq 400$ mrad were recognized by an automated system, the Track Selector [5] [6] and stored on disk. The typical recorded volume was 5 mm $\times$ 5 mm $\times$ 15 mm, centered on the predicted vertex position. This information gave 3-D charged particle tracking segments in each emulsion layer. The 3 coordinates and 2 angles defining each segment were recorded for later analysis. There were approximately $10^4$ track segments found in each emulsion plate.

Tracks that were identified as passing through the volume were used for plate-to-plate alignment and were eliminated as candidate tracks from a neutrino interaction. The distance of closest approach between any two tracks that started in the same or neighboring emulsion plate was calculated and those within 4 $\mu$m were retained as candidates to form a two-track vertex. The data were then searched for two or more sets of two-track vertices being located within 4 $\mu$m of each other. Tracks in these vertex clusters were then fit to a single interaction vertex hypothesis and the tracks were projected into the SFT to find matches, confirming the vertex. A valid vertex was located for 262 of the 499 events. There were three main causes for the location inefficiency: (1) a bias in the charged particle multiplicity of the primary vertex, which reduced the number of events with less than 4 tracks, (2) difficulty in predicting the vertex location for interactions toward the upstream face of each emulsion target due to secondary interactions and (3) uncorrectable alignment problems due to the slipping of a set emulsion plates with respect to another set.

Emulsion plates for located events were rescanned by the automated optical system with a scan volume defined by a transverse area of 2.6 mm $\times$ 2.6 mm centered on the located vertex with a longitudinal dimension of at least 1 cm downstream and 2 plates upstream of the vertex. After further alignment 203 events satisfied the requirement that the track resolution was better than $0.6\mu m$. Reported in this paper are the results of the decay search for these 203 events.

The numbers of $\nu_e$ and $\nu_\mu$ charged current interactions were used as a check of sample bias, and were found to be in agreement with the expected numbers. In the set of 203 events, there are 47 events with an identified muon with momentum greater than 10 GeV/$c$ from the primary vertex. In this $\nu_\mu$ charged current sample, 44±17% of the events were from $\pi$ and $K$ meson decays in the beam dump, not from charm decays. After correcting for the acceptance and the efficiency of the muon detectors, the number of $\nu_\mu$ charged current interactions is 94±17. The number of $\nu_e$ events in the sample was calculated separately using the distribution of energy in the calorimeter. The result obtained is $61 \pm 14$ $\nu_e$ charged current interactions, confirming the expectation of approximately equal number of $\nu_\mu$ and $\nu_e$ events from charm decays. From a set of simulated neutrino events, 47% (95 events) are expected to be $\nu_\mu$ charged current events, 27% (55 events) $\nu_e$ charged current events, 5% (10 events) $\nu_\tau$ charged current events and 21% (43 events) neutral current events.

To perform the decay search, segments from the refined scan were re-aligned and re-linked into tracks. Tracks were then fit to form the interaction vertex. Tracks that originated at the interaction vertex and stopped in the scan volume were candidates for decays. Tracks which originated within one or two plates downstream from the end of the stopped track were designated daughters if they had a distance of closest approach to the parent candidate of less than 10 $\mu$m.

The following selection criteria, which retained 50% of simulated $\tau$ events, were used to identify $\tau$ decay candidates from the kink candidates:

- At least one segment of the parent track is identified in the emulsion data.
- Only one daughter track was associated with a parent track.
- The parent track was $<$ 5 mm long.
- The daughter angle with respect to the parent track was >10 mrad and $<$ 400 mrad.
- The impact parameter of the daughter to the parent track was $<$ 10 $\mu$m.
- The impact parameter of the parent track to the interaction vertex was $<$ 10 $\mu$m.
- The impact parameter of the daughter track to the primary vertex was $<$

500 $\mu$m.

• The daughter track momentum was > 1 GeV/c.

• The transverse momentum of the decay was > 250 MeV/c.

• None of the tracks originating at the primary interaction vertex was identified as a muon or electron.

Four $\tau$ candidates passed the above selection criteria. An additional event satisfied all of the criteria except the last, with an electron track identified coming from the interaction vertex. This event was classified as a charged charm decay from a $\nu_e$ charged current interaction. The emulsion plates for each of these four events were manually rescanned as an additional check. Two plates upstream of the kink were examined using a microscope stage to see if there were any tracks that were not found by the Track Selector.

Applying the efficiencies for the $\tau$ and charm decay selection criteria, including the measured position resolution and the detection efficiency of the emulsion plates, gave an expected number of decays in our sample of 4.2 $\tau$ and 0.9 charm decay, which is consistent with the observation of 4 and 1 events, respectively.

There are potentially two major sources of background to the $\tau$ events. *(i) Charmed mesons* produced in $\nu_e$ or $\nu_\mu$ charged current interactions where the lepton ($e$ or $\mu$) from the primary vertex escapes detection. This background is estimated to be $0.18 \pm 0.03$ events. *(ii) Hadronic secondary interactions* of charged particles from the primary vertex, in which one charged track is found with a kink angle greater than 10 mrad, a decay $p_T$ greater than 250 MeV/$c$, and no lepton observed. From Monte Carlo studies, $0.16 \pm 0.04$ of these events satisfy the selection criteria. Therefore, the total sample background of tau-like events generated by charm or interactions is $0.34 \pm 0.05$. The Poisson probability of the background fluctuating to the signal level is $4.0 \times 10^{-4}$. A description of each $\tau$ event follows, and the events are shown graphically in Fig. 3.

*Event 3024-30175* $\tau \to e + \nu_\tau + \nu_e$: A track from the primary vertex has a kink with an angle of 93 mrad in the 200 $\mu$m plastic base of an emulsion plate, 4.5 mm from the primary vertex. The daughter track was identified as an electron from accompanying electron pair tracks in emulsion data. A shower develops from this track as measured in the SFT confirming the identification. No other tracks were identified as an electron or a muon. The energy was estimated by measuring the multiple scattering of the daughter particle and by the shower development in the SFT. The first method gives $2.9^{+1.5}_{-0.8}$ GeV/c, and is considered a lower limit due to energy loss of the electron in the target material. The latter method gives the value of $6.5 \pm 2.5$ GeV.

*Event 3039-01910* $\tau \to h + \nu_\tau + X$: One track, isolated from the others, has a 90 mrad kink angle, 0.28 mm from the primary vertex. This track is isolated from the other charged tracks and is emitted opposite to the other tracks in the plane perpendicular to the beam. The decay is located in the plastic base (200 $\mu$m thickness). The daughter momentum was measured to be $4.6^{+1.6}_{-0.9}$ GeV/$c$. The estimated decay $p_T$ is $0.41^{+0.14}_{-0.08}$ GeV/$c$. The daughter track was identified as a hadron since a secondary interaction is seen after 1.5 cm.

*Event 3263-25102* $\tau \to h + \nu_\tau + X$: One track has a 130 mrad kink, 1.8 mm from the primary vertex. The decay is located in a steel plate. The daughter momentum is estimated to be $1.9^{+1.6}_{-0.6}$ GeV/$c$. The corresponding $p_T$ is $0.25^{+0.21}_{-0.08}$ GeV/$c$.

*Event 3333-17665* $\tau \to e + \nu_\tau + \nu_e$: The track closest to the beam direction has a 13 mrad kink, 0.54 mm from the primary vertex. The decay is located in the 800 $\mu$m thick plastic base. The daughter is identified as an electron. The momentum of the electron was measured to be $21^{+14}_{-6}$ GeV/$c$ by multiple scattering analysis. Because of the continuous energy loss of the electron, this value must be considered as a lower limit. The corresponding decay $p_T$ limit is $\geq 0.27^{+0.19}_{-0.08}$ GeV/$c$.

It should be noted that since the neutrino flux had only an estimated 5% $\nu_\tau$ component, the possibility that the $\nu_\tau$ is a superposition of $\nu_e$ and $\nu_\mu$ cannot be eliminated using the results of this experiment. Results from other experiments [9] [10] [11], which were sensitive to $\tau$ leptons, show that the direct coupling of $\nu_\mu$ to $\tau$ is very small ($2 \times 10^{-4}$). The upper limit (90% CL) for $\nu_e$ to $\tau$ is much larger, $1.1 \times 10^{-2}$ (90% CL). Assuming this upper limit, the estimated number of $\tau$ events from this hypothetical source is 0.27±0.09 (90% CL).

In summary, in a set of 203 located neutrino interactions, four events have a track that meets all the requirements for $\tau$ decays, with no evidence of another lepton from the primary vertex. The total background is estimated to be $0.34 \pm 0.05$ events. Two events are identified as $\tau \to e\nu_\tau\nu_e$ decays and have negligible level of background from scattering. The probability that the four events are from background sources is $4 \times 10^{-4}$, and we conclude that these events are evidence that $\tau$ neutrino charged current interactions have been observed.

We would like to thank the support staffs at Fermilab and the collaborating institutions. We acknowledge the support from the U.S. Department of Energy, the Japan Society for the Promotion of Science, the Japan-US Cooperative Research Program for High Energy Physics, the Ministry of Education, Science and Culture of Japan, the Korean Research Foundation, and the DOE/OJI Program.

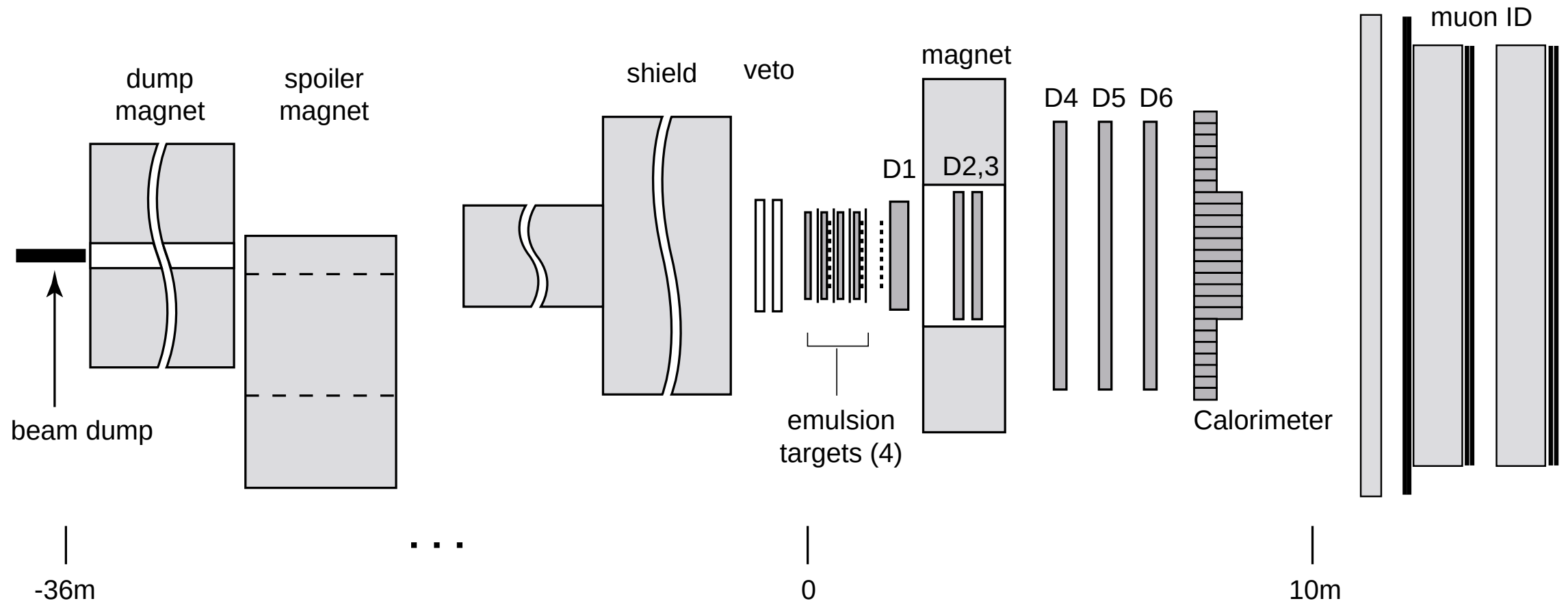

Figure 1: Experimental beam and spectrometer. At the left, 800 GeV protons were incident on the beam dump, which was 36 m from the first emulsion target. Muon identification was done by range in the system at the right.

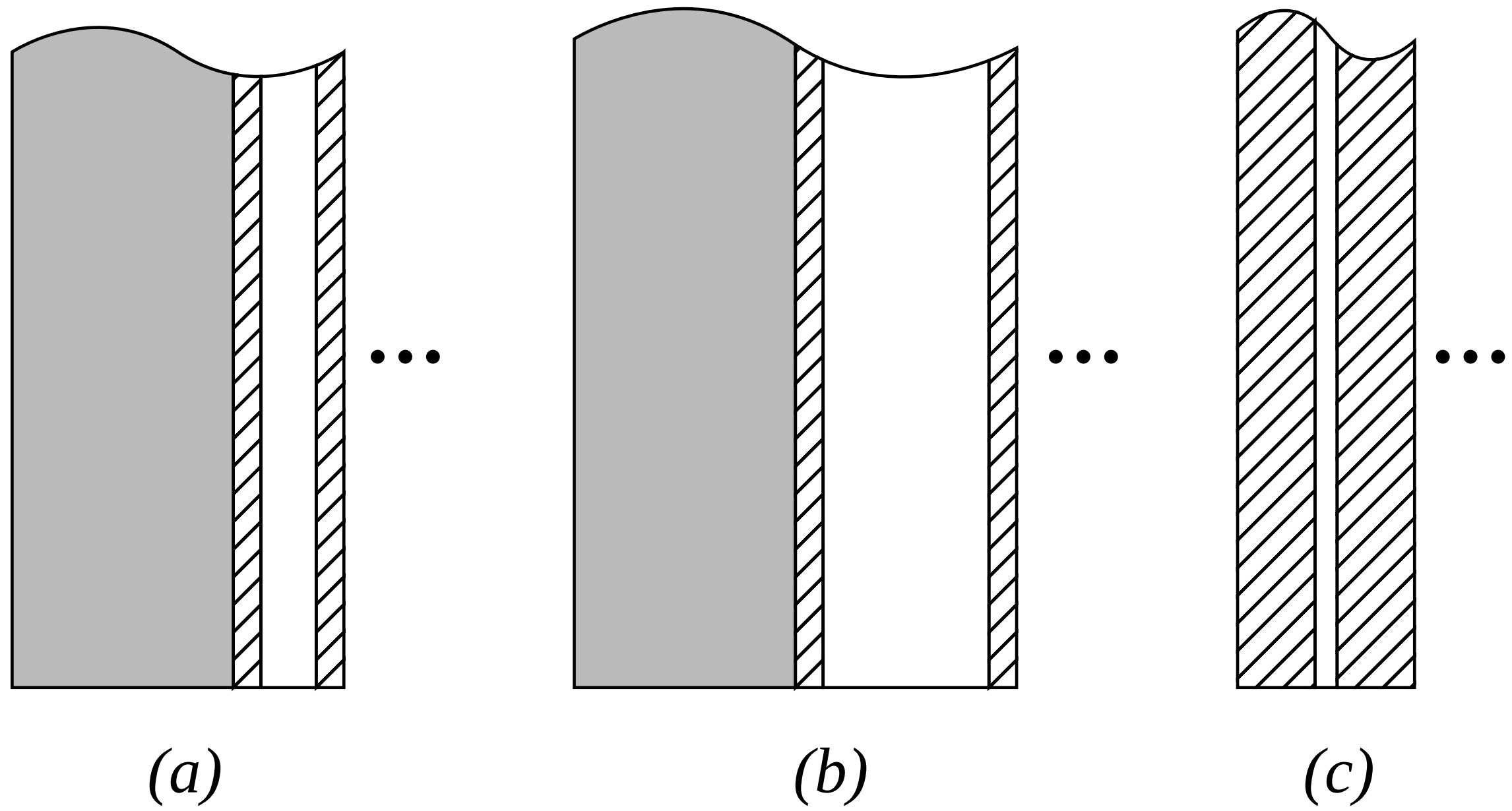

Figure 2: Emulsion target designs. The ECC designs ($a$ and $b$) used stainless steel sheets interleaved with emulsion plates. Most neutrino interactions were in the steel. The bulk emulsion type ($c$) used thicker emulsion layers, without steel. Steel is indicated by shading, emulsion by cross-hatching, the plastic base is unshaded.

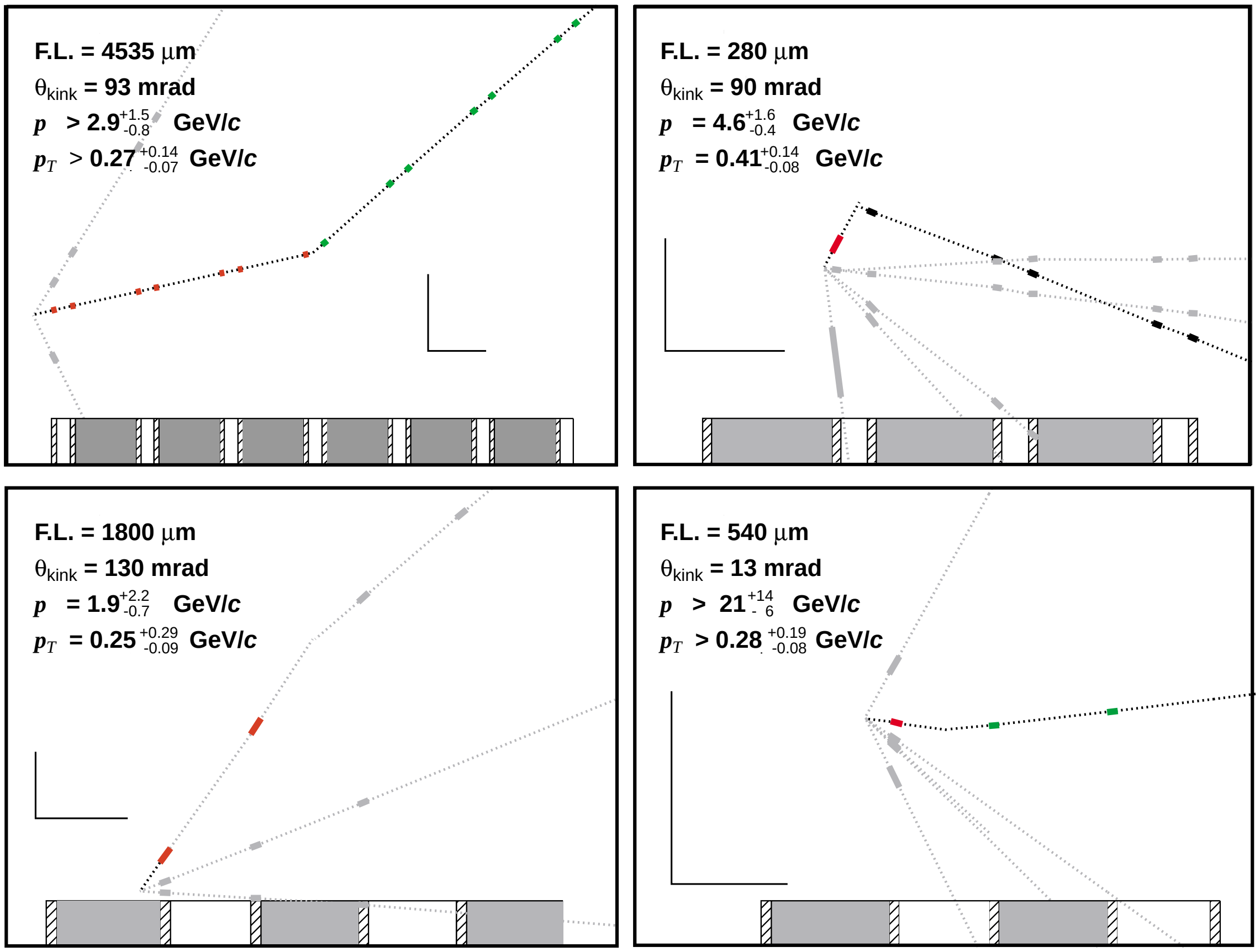

Figure 3: The four $\nu_\tau$ CC interaction events. *(top left)* 3024-30175 *(top right)* 3039-01910, *(bottom left)* 3263-25102, *(bottom right)* 3333-17665. The neutrinos are incident from the left. The scale is given by the perpendicular lines with the vertical line representing 0.1 mm and the horizontal 1.0 mm. The target material is shown by the bar at the bottom of each part of the figure representing steel (shaded), emulsion (cross-hatched) and plastic (no shading).